\relax
\documentclass[letterpaper]{article} %
\usepackage{aaai22}  %
\usepackage{times}  %
\usepackage{helvet}  %
\usepackage{courier}  %
\usepackage[hyphens]{url}  %
\usepackage{graphicx} %
\usepackage{natbib}  %
\usepackage{caption} %
\DeclareCaptionStyle{ruled}{labelfont=normalfont,labelsep=colon,strut=off} %
\usepackage{algorithm}
\usepackage{algorithmic}
\usepackage{color,xcolor}

\usepackage{newfloat}
\usepackage{listings}
\floatstyle{ruled}
\newfloat{listing}{tb}{lst}{}
\floatname{listing}{Listing}

\title{Targeted Data Poisoning Attack on News Recommendation System \\ 
by Content Perturbation}
\author{
    Xudong Zhang,
    Zan Wang,
    Jingke Zhao,
    Lanjun Wang
}

\affiliations{
    School of New Media and Communication, Tianjin University

    Xudongenen@tju.edu.cn, 
    wangzan@tju.edu.cn,
    ke1206371563@gmail.com,
    wanglanjun@tju.edu.cn
}

\usepackage{amsmath}
\usepackage {subcaption}
\usepackage{enumerate}
\usepackage{amsfonts}
\usepackage{multirow}

\newtheorem{problem}{Problem}

\begin{document}

\maketitle

\begin{abstract}
News Recommendation System(NRS) has become a fundamental technology to many online news services.  Meanwhile, several studies show that recommendation systems(RS) are vulnerable to data poisoning attacks, and the attackers have the ability to mislead the system to perform as their desires.  A widely studied attack approach, injecting fake users, can be applied on the NRS when the NRS is treated the same as the other systems whose items are fixed.  However, in the NRS, as each item (i.e. news) is more informative, we propose a novel approach to poison the NRS, which is to perturb contents of some browsed news that results in the manipulation of the rank of the target news. Intuitively, an attack is useless if it is highly likely to be caught, i.e., exposed. To address this, we introduce a notion of the exposure risk and propose a novel problem of attacking a history news dataset by means of perturbations where the goal is to maximize the manipulation of the target news’ rank while keeping the risk of exposure under a given budget.  We design a reinforcement learning framework, called TDP-CP, which contains a two-stage hierarchical model to reduce the searching space.  Meanwhile, influence estimation is also applied to save the time on retraining the NRS for rewards.  We test the performance of TDP-CP under three NRSs and on different target news.  Our experiments show that TDP-CP can increase the rank of the target news successfully with a limited exposure budget.  
\end{abstract}

\section{Introduction}

Online news services such as Google News and Microsoft News have become important platforms for a large population of users to obtain news information~\cite{das2007google,wu2020mind}. Massive news articles are generated and posted online every day, making it difficult for users to find interesting news quickly. The recommendation system is playing an increasingly important role in online news service platforms. It provides a data-driven approach to predict online users’ potential preferences on the latest news by learning their historical clicking behaviors.  

The classical methods are collaborative filtering(CF)~\cite{rendle2010factorization} and Logistic Regression(LR)~\cite{montanes2009collaborative}, which can help us make predictions about users’ potential interests by collecting other users’ preference information. In recent years, many news recommendation algorithms based on deep neural networks have been proposed and achieved great success in practical applications~\cite{guo2017deepfm,wu2019neural,wu2019npa,wu2019neural-2,an2019neural}.

There have been some evidences showing that existing recommendation systems are vulnerable under data poisoning attacks~\cite{Chen2019Data,zhang2020practical,PoisonRec,lin2020attacking,fang2020influence}.  As almost all of these studies focus on general recommendation scenarios, such as products, movies, books, etc.  The attack approaches can be roughly divided into two types: fake rating scores and fake users.  For example, \citet{Chen2019Data} formulate the task of poisoning a cross-domain recommendation system as a bi-level optimization problem and provide the solution on an optimized user-item rating matrix of the target domain; meanwhile, \citet{PoisonRec} propose an adaptive data poisoning framework by actively injecting fake users into the recommendation system to increase exposure of a target item set.  

For news recommendation systems, we can inject fake users to achieve the goal of poisoning. However, some recent studies propose defense methods, such as detect fake users~\cite{kumar2015ensemble,aktukmak2019quick} and adversarial train~\cite{wu2021triple,wu2021fight}, which mitigate the damage of poisoning.  Nevertheless, news recommendation is different from other scenarios, because the item for recommendation is news, which contains more information and provides more spaces to be perturbed.  As a result,  in this study, we propose a novel attack approach for news recommendation systems which is to \textit{perturb the viewed news} for target news promotion. 

Since to perturb one sample in the viewed dataset influences several user historical behaviors, it is trivial to replace some high frequency viewed news with the target news, which elaborates the prediction of clicking target news for other users.  However, as news is timeliness, replacing the whole article with the latest content is easy to be detected.  Therefore, we specify the novel attack approach as to \textit{perturb the content of the viewed news in an imperceptible manner}.  

To our best knowledge, it is still an under-explored problem to learn attack strategies with respect to news content perturbation. Furthermore, in order to make the attacking scenario more practical, we set the online recommendation system as a black box, whose parameters and model architectures are unknown to the attacker. As a result, there are at least three key challenges to achieve this new attack.  Firstly, the attacking method needs to be generic for different recommendation algorithms.  Secondly, the search space is huge because the perturbation can be applied in the granularity of words. Lastly, since it is impossible to know when the system to be retrained with the perturbation, the feedback to the attacking mechanism takes a long time.

To tackle these challenges, we first formally introduce the novel problem of targeted data poisoning attack under an exposure risk budget on news recommendation systems, as a constrained optimization problem. Then, we show that finding a sequence of perturbations can be naturally modeled as Markov Decision Process (MDP)~\cite{PoisonRec,zhang2020practical,9458733}. We propose a solution of Targeted Data Poisoning on news recommendation with Content Perturbation(TDP-CP) framework based on reinforcement learning(RL), leveraging the fact that RL algorithms using policy networks such as Deep Q Network(DQN) to solve MDPs \cite{mnih2016asynchronous}. To deal with the difficulty of the huge search space, we propose a two-stage hierarchical structure to 1) decompose news-level and content-level of searching; 2) leverage a complete binary tree to shrink the space to those with close effectiveness on attacks. Finally, to solve the bottleneck on obtaining the reward by retraining the recommendation model, an influence estimator is designed to estimate the reward of perturbations as the feedback of the RL system.

To summarize, the major contributions are as follows:
\begin{enumerate}
    \item We propose a novel data poisoning approach for a news recommendation system of perturbing the news content. We then formally define it as a constrained optimization problem and model the approach to find the solution as MDP.  
    \item An RL framework, TDP-CP, is proposed with a DQN to obtain the perturbations. In order to reduce the search space, we design a two-stage hierarchical structure.   
    \item We also implement an influence estimator to obtain the reward without retraining the model.
    \item Our experiments show that TDP-CP can increase the rank of the target news successfully with a limited exposure budget.  
\end{enumerate}

\section{Related Work}
\subsection{News Recommendation System} The classical news recommendation system is mainly adopted Logistic Regression(LR) and Factorization machine(FM). LR recommends news that is directly related to user's behavior, but it has lower generalization and needed manual feature engineering\cite{montanes2009collaborative}. On the contrary, FM has higher generation and less human participation, but it is hard to learn effective embedding when the amount of data is little\cite{rendle2010factorization}. 

In recent years, several deep learning based news recommendation methods are proposed. \citet{cheng2016wide} proposed Wide \& Deep method by combining LR and deep neural network(DNN). \citet{guo2017deepfm} proposed the DeepFM method by combining FM and DNN, where FM extracts low-order features while DNN extracts high-order features. Meanwhile, DeepFM has a shared input to its ``wide'' and ``deep'' parts, with no need for feature engineering besides raw features. 

Later, the attention mechanism becomes to be applied on news recommendation as the item for recommending containing more information.  \citet{wu2019neural} proposed NAML which can learn informative representations of users and news by exploiting different kinds of news information base on self-attention mechanism. \citet{wu2019neural-2} proposed NRMS which uses multi-head self-attention to learn news representations from news content and user representations from browsed histories. 

Along with pre-trained language models, \citet{sun2019bert4rec} proposed BERT4Rec method. The core idea of this method is that it employs the deep bidirectional self-attention to model user behavior sequences. To avoid the information leakage and efficiently train the bidirectional model, it also adopts the Cloze objective to a sequential recommendation. In Sec.~\ref{sec:exp}, we apply DeepFM, NAML and NRMS to evaluate the performance of our attack.

\subsection{Data Poisoning on Recommendation System}\label{subsec:dp}
There has been extensive prior research concerning the data poisoning issues on recommendation systems from classical FM~\cite{li2016data,Chen2020data} to 
DNN~\cite{Chen2019Data,zhang2020practical,PoisonRec,lin2020attacking,fang2020influence}. Almost all of these studies are designed for general recommendation scenarios, such as products, movies, games, etc. So far, we find only one recent study \cite{descampe2021automated} discussed the specific poisoning issue on news recommendation systems.  However, \citet{descampe2021automated} focused on the algorithmic accountability, but did not provide a technical solution on attacking or defending. 

For those studies on general recommendation systems, two attack approaches are widely applied. One is to inject fake rating scores of items~\cite{li2016data,Chen2019Data,Chen2020data}, which can not be applied to news recommendation system, because of different settings with respect to the scenario. The other type is to inject fake users~\cite{fang2020influence,PoisonRec,zhang2020practical,huang2021data}, which we can applied on news recommendation systems. 

However, a lot of defense approaches against injecting fake users also have been proposed, including attack detection~\cite{kumar2015ensemble,aktukmak2019quick} and adversarial training~\cite{wu2021triple,wu2021fight}.  By applying these defense methods can mitigate the issues from fake users.  

Nevertheless, in the domain of news recommendation system, since each recommending item, i.e. news, contains much more information, which indicates more spaces to be perturbed.  As a result, this study proposes a novel attack approach, which is to perturb the visited news in an imperceptible manner to achieve the poisoning goal.    

\section{Problem Formulation}\label{sec:prob}
In this study, we focus on the news recommendation task.  
Assume there is a set of users $\mathcal{U} = \{u_1,u_2,...,u_M\} $, and a click history $\mathcal{A}_m$ of each user $u_m \in \mathcal{U}$ which is a subset of the full viewed article set $\mathcal{A}^r$, i.e., $\mathcal{A}^r=\cup_{m=1}^M \mathcal{A}_m$. The recommendation model $\mathbf{F}$ is trained based on the set of click histories $\mathbb{A}=\{\mathcal{A}_1,\ldots,\mathcal{A}_M\}$.  In the inference stage, given a candidate news $o$,  $\mathbf{F}$ takes a specific user's history as input and calculates the probability of the user clicking the candidate news $o$, denoted as $\mathbf{F}(o)$.
Considering a set of candidate news $\mathcal{A}^c$, a ranked list can be generated based on probabilities. Denote $k_{m}^{o}$ as the rank of the news $o$ in the recommended list of the user $u_{m}$.

Given a recommendation system, the goal of the attacker is to promote target news to a higher rank in the recommendation list in as many users as possible.  We denote $V(\cdot): \mathcal{A}^c \rightarrow \mathbb{R}$ the \textbf{\textit{rank score}} that maps a candidate news to its ranking quality of the given recommendation system among users.  There are several widely used metrics to evaluate the ranking quality~\cite{chen2017performance}.  For example, mean reciprocal rank (MRR), which is the harmonic mean of ranks of the target news among all users; mean average precision (MAP), which is the average precision of recommendation lists of all users by considering the target news as the hit. 

The rank score function $V(\cdot)$ is uniquely determined by the recommendation model trained by $\{\mathcal{A}_1,\ldots,\mathcal{A}_M\}$.  In this study, we use $V_{\mathbb{A}}(o^*)$ to represent the ranking quality with respect to the target news $o^*$. If $\mathbb{A}$ is deliberately perturbed carefully, the rank score function can be altered to manipulate the rank of the candidate set.  Such a manipulation of the rank score of given target news is called a \textbf{\textit{targeted data poisoning attack}}~\cite{li2016data}, which usually consists of a sequence of perturbations.  

Note that an attacker can also demote the target news.  Similar to other data poisoning attacks~\cite{zhang2020practical}, we regard demotion can be viewed as a special case of promotion as an attacker can promote the other news in $\mathcal{A}^c$ to achieve the target news is demoted in the list.  As a result, we focus on promoting the target news as the attack.

To achieve the attack goal, we propose a novel scenario in which the attacker can perturb the news from the viewed article set $\mathcal{A}^r$. When $\mathcal{A}^r$ is perturbed, $\mathbb{A}$ is also changed correspondingly, because $\forall o^r \in \mathcal{A}^r$, it exists at least one of $\mathcal{A}_m$, $m=1,\ldots,M$. As a result, $\mathbf{F}$ is changed to achieve the attacker's goal after retrained.   

The \textbf{\textit{perturbation}} is an operator applying on the viewed news, denoted as $\pi(o^r)$.  As news is informative, there is a large space of the operators from the article level to the word level.  
For a sequence of perturbations $\Pi(\mathcal{A}^r) =[\pi(o^r_1),\ldots,\pi(o^r_T)]$ is uniquely determined by permutation of news in $\mathcal{A}^r$ and combination with operators.

Meanwhile, depending on the history news $o^r \in \mathcal{A}^r$ chosen to be modified as well as different operators, different perturbations have different risks of being detected.  We model the risk of a perturbation by an \textbf{\textit{exposure risk}} function $C(\cdot): (\mathcal{A}^r, \Pi) \rightarrow \mathbb{R}$ that maps the news $o^r$ to the risk score $C(\pi(o^r))$ with a modification operator $\pi \in \Pi$.   A higher value of $C(\pi(o^r))$ means the perturbation on $o^r$ is easy to be detected.  For  $\Pi(\mathcal{A}^r)$,  the risk function to evaluate the exposure risk of perturbing $\mathcal{A}^r$ with $\Pi$ is denoted by $C(\Pi(\mathcal{A}^r))$.

We formally state our problem of \textbf\textit{{targeted data poisoning with a risk constraint}} as: 
\begin{problem}
Given a recommendation system $\mathbf{F}$ trained by a set of click histories $\{\mathcal{A}_1,\ldots,\mathcal{A}_M\}$, target candidate news $o^*$, a rank score function $V_{\mathcal{A}^r}(o^*)$ where $\mathcal{A}^r=\cup_{m=1}^M \mathcal{A}_m$, a risk exposure function $C(\cdot)$, and a risk budget $\bar{C}$, the problem of risk-constrained targeted data poisoning problem is to find a sequence of perturbations $\Pi(\mathcal{A}^r)$, such that $V(o^*)$ is maximized while keeping the exposure risk bounded, i.e. $C(\Pi(\mathcal{A}^r))\leq \bar{C}$.  The optimization problem is denoted as:
\begin{equation}
    \begin{split}
        \arg\max_{\Pi(\mathcal{A}^r)} V_{\Pi(\mathcal{A}^r)}(o^*) \\
        s.t.\ C(\Pi(\mathcal{A}^r)) \leq \bar{C}
    \end{split}
\end{equation}
\end{problem}

In this paper, we assume that the attacker is granted the following \textit{knowledge and capability}, which is following the previous study \cite{zhang2020practical} but we tailor unnecessary ones to fit for our attack approach.
\begin{enumerate}
    \item The attacker can access the full activity history of all the users in the news recommendation system.
    \item The recommendation system is black-box to the attacker, i.e., the recommendation model, including parameters and the architecture, is unknown.
    \item The attacker does not know when the targeted black-box recommendation model is retrained.  
\end{enumerate}

\section{Methodology}
\subsection{Framework}
Our problem of risk-constrained targeted data poisoning can be modeled as Markov Decision Process(MDP).  In this study, the states of MDP correspond to the full viewed set after perturbed.  Irrespective of the order in which news will be perturbed, the next choice only depends on the current state.  For training MDP, Reinforcement Learning (RL) is an effective way because RL can learn a series of actions in an unknown environment using a trial and error method.

\begin{figure}[t]  
\centering
\includegraphics[width=0.45\textwidth]{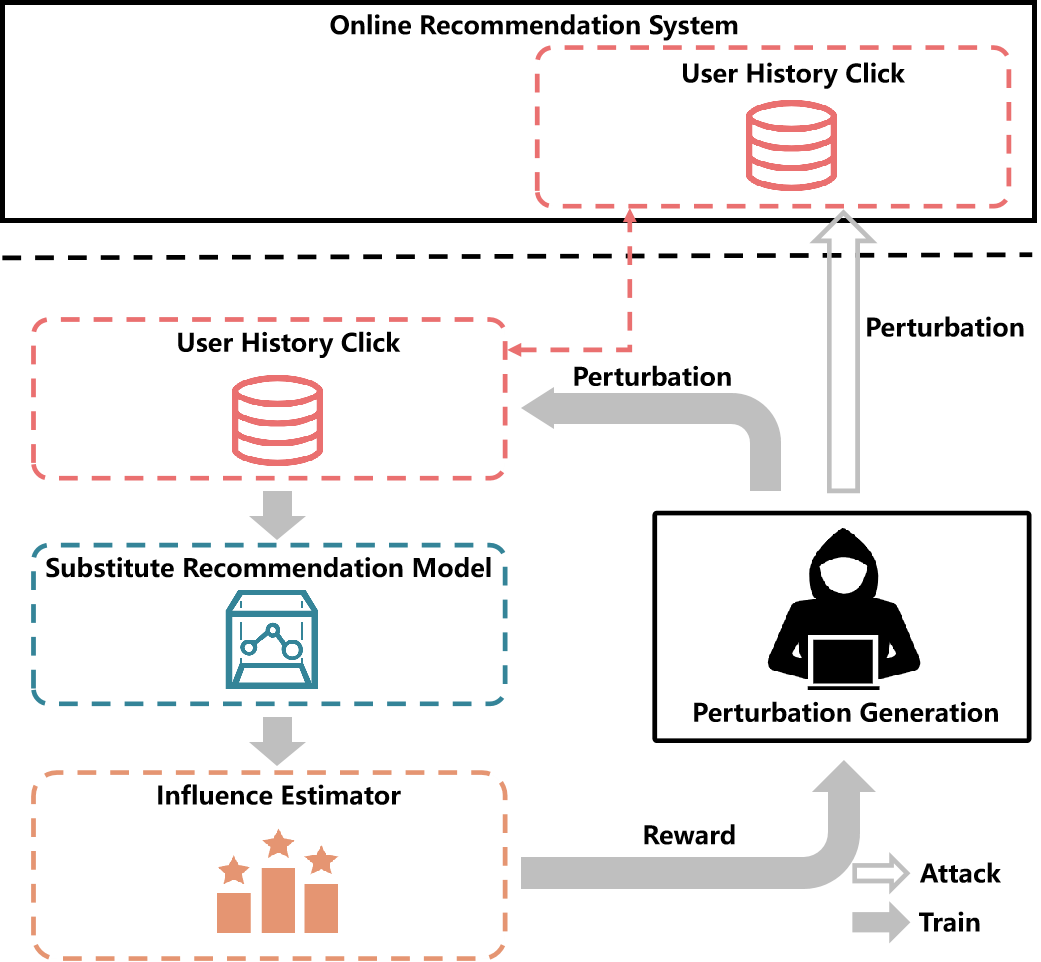}
\caption{Overview of TDP-CP Framework.}\label{fig:framework}
\label{fig:Overview}
\end{figure}

We design a framework as shown in Fig.~\ref{fig:framework}, Target Data Poisoning attack with Content Perturbation(TDP-CP). The main component of this framework is \textit{Perturbation Generation Environment}, which is to generate the perturbed history news by RL. As the \textit{online news recommendation system} is a black-box for attackers, an \textit{offline substitute system} is set to provide the reward for the generator.  In order to obtain the reward, one straightforward approach is to retrain the offline model, however, it costs a lot of time and requires more resources.  To improve the efficiency of obtaining the reward, we design an \textit{Influence Estimator} based on the offline substitute system. We will introduce detailed designs in the rest of this section.

\subsection{Perturbation Generation Environment}
MDP is defined as $\mathcal{M}(\mathcal{S}, \mathcal{N}, \mathcal{P}, R)$,  where $\mathcal{S}$ is a set of states, $\mathcal{N}$ is a set of actions,  $\mathcal{P}$ is the transition probabilities governing the probability of moving from one state to another, $R$ denotes the reward of performing an action at a given state.  A path is a sequence of the states that are visited.  For a finite horizon MDP, every possible path has a finite length.  A terminal state is the last state at the end of a path.  
The detailed model of the MDP is as follows.

\textbf{State} $\mathcal{S}$ is the set of all states.  At a given time step $t$, the corresponding state $s_t \in \mathcal{S}$ corresponding to the full viewed set after perturbed at time $t$, denoted by $\Pi(\mathcal{A}_t^r)$, and the remaining budget available, $C_t$.  When $t=0$, $C_0=\bar{C}$, and $\mathcal{A}_0^r = \mathcal{A}^r$, the initial state is the full viewed set, prior to any attack. At every time step $t$, an operator $\pi_t$ is applied on a selected news $o_t^r$, Meanwhile, the remaining risk budget is also update follows: $C_{t} = C_{t-1} - C(\pi(o^r_t))$.  Finally, we model a state $s_t$ as the set of embedding of the perturbed news along-with the available budget $C_{t}$ at that state.

\textbf{Action}  In general, we can roughly divide the action into two stages.  The first stage is to select the news from $\mathcal{A}^r$ to perturb, and the second stage is to perturb the content of the news.  There are various ways to perturb the content, such as replacing one word with one from the target, generating new content based on the target, etc. As an action has two stages, we can specify the risk function as: 
\begin{equation}\label{eqn:grisk}
C(\pi(o^r)) = freq(o^r) + D(o^r, \pi(o^r))
\end{equation}
where $freq(\cdot):\ \mathcal{A}^r \rightarrow [0,1]$ is the frequency of the selected article $o^r$ clicked by users in $\mathcal{U}$; $D(o^r, \pi(o^r)): (\mathcal{A}^r, \Pi) \rightarrow [0,1]$ denotes the distance between the original news and the perturbed news. 

Generally, the search space is  $\mathrm{O}(|\mathcal{A}^r|\times \mathrm{N}) $, where $\mathrm{N}$ denotes the search space of the perturbation in the content level.  In this study, we design a hierarchical structure to reduce the search space, similar to what is used in \citet{takanobu2019aggregating,PoisonRec}, but extend it to two stages.  This two-stage hierarchical structure reduces the search complexity to $\mathrm{O}(\log|\mathcal{A}^r| + \log \mathrm{N})$.  Fig.~\ref{fig:hs} demonstrates the structure of the two-stage hierarchical structure.  We will discuss it in detail in Sec.~\ref{subsec:opt}.

\textbf{Reward} We set the reward is directly inline with our optimization objective, which is to increase the rank score of the target news.  
Meanwhile, a discount factor $\gamma$ intuitively regulates the amount of future information that the agent can foresee. We set it as $\gamma = 0.99$ and the rank score is calculated after the complete attack path is injected into the offline system.  

\textbf{Termination} The termination condition is governed by the risk budget $\bar{C}$.  An MDP path is terminated when the remaining budget $C_t$ at any state $t$ is not enough to select any further action.  Besides, we enforce that MDP has a finite horizon, hence if the budget does not exhaust after $T$ steps, we terminate the process.  $T$ is set as a hyper-parameter which denotes the maximum possible length of a path.

There are different optimization methods used to train an RL agent to generate perturbations, such as Q-learning \cite{watkins1989learning}  and  Advantage  Actor-Critic(AAC) \cite{mnih2016asynchronous}. Q-learning, in particular, is shown to perform well under finite-horizon  MDPs \cite{dai2018adversarial}. Besides, as the non-linearity of the recommendation system, we apply deep Q networks(DQN) to model the Q value of each action. 

\subsection{Optimization}\label{subsec:opt}
In this paper, we use a DNN to generate the embedding and long short-term memory(LSTM) to generate the optimal perturbation sequence, which is a widely used approach in DQN~\cite{hausknecht2015deep}. More importantly, we design a two-stage hierarchical structure to reduce the action space.

The idea of the hierarchical structure is based on an assumption that close effectiveness of perturbations has similar attack properties in a recommendation system both at article level and content level. That is to say, for a sampled perturbation sequence, perturbing news or content with similar ones will not obviously affect its poisoning performance.  Here, we use effectiveness to evaluate the performance of the perturbation.  In the article level, as higher frequency articles perturb the training data more, but much closer to the target makes a fewer contribution to the rank score improvement, as a result, we define the effectiveness as $\frac{freq(o^r)}{D(o^r, o^*)}$.  The effectiveness of the content level can be set in the similar way.  Take the word replacement as an example, the effectiveness in the content level is as $\frac{freq(w_i^r)}{D(w_i^r, w_j^*)}$, where $freq(w^r_i)$ is the frequency of the word $w_i$ in the news $o^r$, $w_j^*$ is a word from the target news, and $D(w_i^r, w_j^*)$ is the distance between two words.

In the following, we introduce the construction of the two-stage hierarchical structure and how to sample the action.

\begin{figure}[h]
    \centering
    \includegraphics[width=8cm]{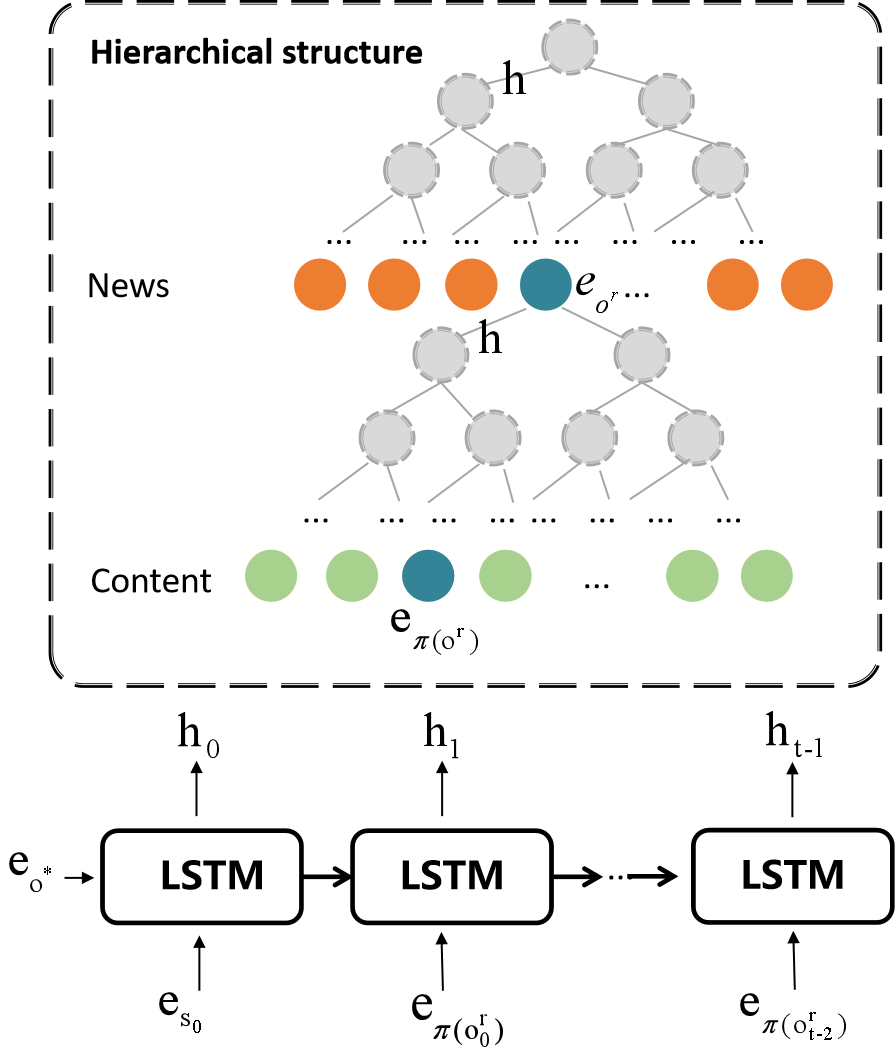}
    \caption{Two-Stage Hierarchical Structure.}
    \label{fig:hs}
\end{figure}

\textbf{Construction of the news level.} As shown in the Fig.\ref{fig:hs}, the top stage of the two-stage hierarchical structure is the selection of the perturbation news. we adopt a complete binary tree for the news level hierarchical structure. More specifically, we introduce the construction of the complete binary tree for history browsed news in detail. The tree each layer except the last has 2$d$ nodes ($d$ is the depth of the current layer). Furthermore, the nodes at the last layer are required to be left-aligned with their effectiveness value. When building this complete binary tree, we need to repeatedly insert nodes from left to right in one layer of the tree. If a layer is full, we insert nodes into the next new layer. The building process is stopped when the number of leaf nodes is equal to $|\mathcal{A}^r|$. All non-leaf nodes are simulated nodes that are used to assist the sampling process on the tree. For leaf nodes, we assign the news embedding on $\mathcal{A}^r$, where $\forall o^r \in \mathcal{A}^r$, there is $e_{o^r}$. Since only leaf nodes, which are real history news, have their own features, we randomly initialize trainable feature representations for all simulated non-leaf nodes, whose dimension is equal to that of real news. 

\textbf{Construction of the content level.} The hierarchical structure of the content level is extremely similar to the news level. The root node of the content level complete binary tree is the leaf node where the selected modified news is located.

\textbf{Sample on the hierarchical structure.}  Fig.\ref{fig:hs} shows the structure of the policy network. With a built hierarchical structure, we can sample news following the tree structure more efficiently. First, we obtain the hidden layer $h_0$ from the first LSTM output by given the target news $o^*$ and initial state, i.e. $ h_{0}=\operatorname{LSTM}\left(e_{o^*},e_{s_0}\right)$, which will be used to pick the first action $\pi(o^r_1)$. Next, we perform how to select the news.  We need to sample a complete path on the news level complete binary tree from the root to the leaf.  Given a layer $d$,  we select the left or right node by comparing their $\exp \{\mathbf{D}(h_t)\cdot e_{node}\}$, where $e_{node}$ is the embedding of the node, and $\mathbf{D}$ is the DNN to make the output dimension as equal to $|e_{node}|$. After determined the leaf node $o^r_1$, the content level binary tree is also selected in the same manner.  Finally, we can obtain the action $\pi(o^r_1)$ and calculate the remaining risk budget correspondingly. This process runs for $T$ times or terminates until the risk budget $\bar{C}$ is used out.

\subsection{Influence Estimation}\label{subsec:infest}
We need to utilize the manipulation outcome of the current perturbation samples as reward feedback to update the policy network. The most straightforward way to obtain the outcome is to retrain the entire model. However, retraining the deep news recommendation system is prohibitively slow. To make the attack methodology practical, we propose to use influence function for an efficient estimation of the manipulation outcome. 

We set up a general scenario that the offline recommendation system consists of multiple recommendation models, which are trained on the same dataset. Recommendation results from these models are aggregated via weighted scores. That is, suppose $Q$ different recommendation models are deployed to recommend news for user $u$ and the score of news $o$ in the $q$-th model is denoted as $F_q(o)$. Here, we define the score of the news are obtained as follows:
\begin{equation}\label{eqn:offline}
\mathbf{F}(o)=\frac{1}{Q} \sum_{q=1}^{Q} w_{q} \cdot \mathbf{F}_{q}(o) 
\end{equation}
where $w_q$ stands for the weight of the $q$-th recommendation model. Ideally, these weights are used to adjust the offline substitute system to match the characteristics of the online recommendation model via comparing the recommendation lists.

With the offline system, we can implement the estimation of the influence of the perturbed samples. Formally speaking, the parameter estimator of the recommendation models on the clean dataset is:
\begin{equation}
\hat{\theta}:=\arg \min _{\theta} \frac{1}{|\mathcal{U}|} \sum_{u \in \mathcal{U}} \mathcal{L}\left(\mathcal{A}_u ; \theta\right)
\end{equation}
where $\theta$ denotes the parameter vector, $\mathcal{L}$ stands for the loss function of the recommendation model. 

Suppose we perturb a sample $o^r$ with a perturbation $\delta$ to upweight a small $\epsilon$ of the loss, and $\mathcal{U}^\prime$ denotes the set of users whose historical records contain $o^r$, the estimation of $\theta$ with respect to $\delta$ is as:
\begin{equation}
\hat{\theta}_{\delta}:=\arg \min _{\theta} \frac{1}{|\mathcal{U}|} \sum_{u \in \mathcal{U}} \mathcal{L}\left(\mathcal{A}_u ; \theta\right) + \epsilon  \sum_{u \in \mathcal{U} ^ \prime} \mathcal{L}\left(\mathcal{A}_u  ; \theta\right)
\end{equation}
When $\epsilon \rightarrow 0$, according to the classic results in \cite{cook1982residuals}, the
influence of $\delta$ on the parameter $\theta$ is given by:
\begin{equation}
\frac{d \hat{\theta}_{\delta}}{d \epsilon}=\hat{\theta}_{\delta}-\hat{\theta} \approx-H_{\hat{\theta}}^{-1}  \sum_{u \in \mathcal{U} ^ \prime} \nabla_{\theta} \mathcal{L}\left(\mathcal{A}_u  ; \hat{\theta}\right)
\end{equation}
where $H_{\hat{\theta}}= \sum_{u \in \mathcal{U}} \nabla_{\theta}^{2} \mathcal{L}\left(\mathcal{A}_u, \theta\right)$ is the Hessian matrix of the loss function. 
Here, the key computation bottleneck lies in the calculation of the huge inverse Hessian matrix $H_\theta^{-1}$. Given a sample $o^r$, we use implicit Hessian-vector products~\cite{koh2017understanding} to efficiently
approximate $-H_{\theta}^{-1} \sum_{u \in \mathcal{U} ^ \prime} \nabla_{\theta} \mathcal{L}\left(\mathcal{A}_u  ; \hat{\theta}\right)$.

Based on an approximate estimate of the sample increasing influence on parameter $\hat{\theta}$, we further calculate the influence on the prediction probability function w.r.t. the perturbation. Specifically, suppose we want to promote the target news $o^*$ to more users, all the rank scores rely on the prediction click probability of $o^*$, which is influenced by the perturb of $o^r$ as
\begin{equation}
\begin{aligned}
\frac{d \mathbf{F}\left(o^* ; \hat{\theta}\right)}{d \epsilon}  &  =     \frac{d \mathbf{F}\left(o^*; \hat{\theta}\right)}{d \hat{\theta}_{\delta}} \cdot \frac{d \hat{\theta}_{\delta}}{d \epsilon} \\
& \approx-\nabla_{\theta} \mathbf{F}\left(o^* ; \hat{\theta}\right) H_{\hat{\theta}}^{-1} \sum_{u \in \mathcal{U} ^ \prime} \nabla_{\theta} \mathcal{L}\left(\mathcal{A}_u  ; \hat{\theta}\right)
\end{aligned}
\end{equation}
where $\nabla_{\theta} \mathbf{F}\left(o^* ; \hat{\theta}\right)$ can be obtained by the offline substitute recommendation models.  
\section{Experiments}\label{sec:exp}
In this section, we evaluate the proposed TDP-CP on real against different deep news recommendation methods.  We describe the experimental setup and analyze the performance of the compared baselines and key components.  More experiment results will be described in Appendix. 

\subsection{Experimental Setup}
\subsubsection{Dataset}
Our experiments are conducted on a recently published news recommendation benchmark dataset, MIND~\cite{wu2020mind}. We randomly select 500 news in the validation set of MIND as the target news and divided them into the training set, validation set, and test set in the ratio of 7:1.5:1.5. We use the training set of MIND as click histories to perturb and leverage the rest of the validation set of MIND to evaluate the promotion of the targets. The test set of MIND is ignored in this study because it does not have any clicking annotation.

\subsubsection{Recommendation System.} In the evaluation, we choose five different widely-used representative deep news recommendation algorithms.  Three of them, DeepFM \cite{guo2017deepfm}, NAML \cite{wu2019neural} and NRMS \cite{wu2019neural-2}  to provide online services, while the rest two, LSTUR \cite{an2019neural} and NPA \cite{wu2019npa} consist of the offline substitute system for reward calculation.  To make the system simple, we set the weight of LSTUR and NPA as 0.5 in Eq.~\ref{eqn:offline}.

\subsubsection{Rank Score Function} As discussed in Sec.~\ref{sec:prob}, we set the rank score function used in the evaluation as mean reciprocal rank (MRR), which denotes the harmonic mean of ranks of the target news among all users.  The MRR is defined as:
\begin{equation}
    V_{\mathcal{A}^r}(o^*) = \frac{1}{M} \sum_{m=1}^M \frac{1} {k_m^{o^*}}
\end{equation}
where $k_m^{o^*}$ is the rank of target news $o^*$ in the recommended result list for the user $u_{m}$. 

\subsubsection{Risk Function}  As shown in Eq.~\ref{eqn:grisk}, the risk function depends on the frequency of the selected news and the operation on the perturbed content.  In the evaluation, we set up the perturbation operator as replacing one word $w_i$ from the title of the selected news with one word $w^*_j$ from the title of the target news.  Thus, we set up the risk function used in the evaluation as:
\begin{equation}\label{eqn:risk}
C(\pi(o^r)) = freq(o^r) + D(w^r_i, w^*_j)
\end{equation}
where $D(w^r_i, w^*_j)$ is the cosine distance between replacement words with respect to the embedding from GloVe~\cite{pennington2014glove}.  Note that the cosine distance is normalized to [0,1] by dividing 2.

\subsubsection{Baselines.} As discussed in Sec.~\ref{subsec:dp}, there is no existing work solving exactly the same attack approach. 
Hence, we compare the proposed TDP-CP with several heuristic attack strategies.
\begin{itemize}
\item \textit{None:} No attack is conducted.
\item \textit{Random:}  The attacker randomly selects the news and the content to perturb.
\item \textit{Effective:} The attacker weighs the popularity of different news against the cost of changes, i.e. $\frac{freq(o^r)}{D(o^r, o^*)}$, and chooses the most effectiveness news to modify; then again apply the content based effectiveness, i.e. $\frac{freq(w^r_i)}{D(w^r_i, w^*_j)}$, and choose the most effectiveness word replacement pair. 
\end{itemize}

\subsubsection{Evaluation Metric.} We use the MRR value of the target news, which denotes the rank of target news in the candidate set, as our evaluation metric. The larger MRR is, the better the attack approach performs. The average MRR over 500 targeted news as well as the standard deviation are reported.

\subsubsection{Hyper-parameters.} A two-layer fully connected network is applied to obtain the embedding of hidden states.  The size of each layer is 256, and the activation function is ReLU. We use ADAM~\citep{kingma2015adam} as the optimizer with a learning rate $\alpha =0.001$.  The total number of episodes of RL training is 2000.

\begin{figure}[h]
    \centering
    \includegraphics[width=8cm]{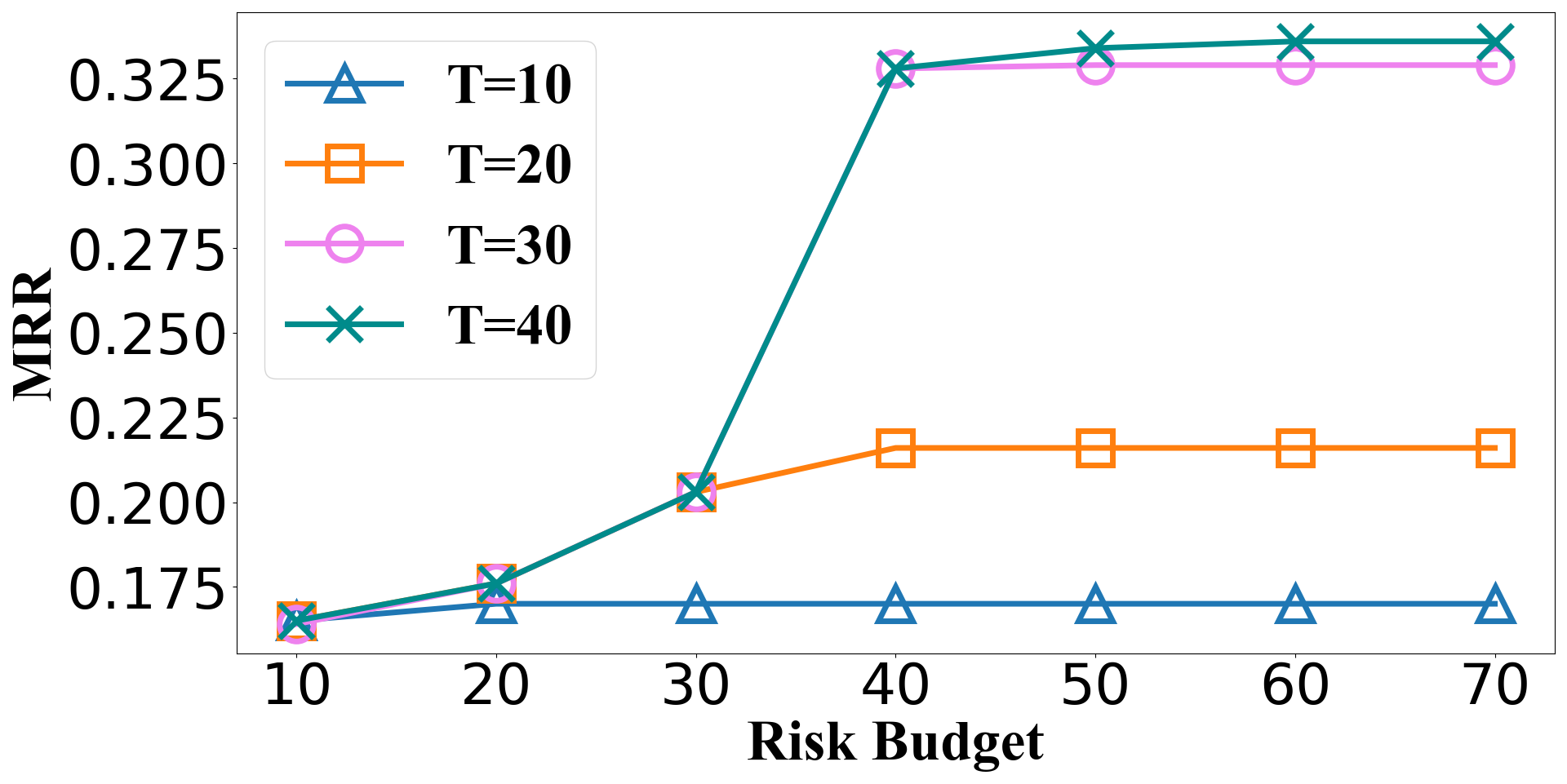}
    \caption{Hyper-parameters Analysis.}
    \label{fig:BT}
\end{figure}

Furthermore, two important hyper-parameters in our task are the risk budget $\bar{C}$ and the maximum number of perturbations $T$. We evaluate how these two hyper-parameters influence MRR on NRMS (note that the other two perform similarly). As shown in the Fig.~\ref{fig:BT}, when $T$ is relatively small, the attack effect is mainly limited by it. As $T$ becomes larger, the budget $\bar{C}$ becomes the main constraint. According to the figure, due to $\bar{C}=40$ and $T=30$ bring a big jump of the rank score value, we use them in the following experiments if there is no further specification.

\subsection{Attack Performance}
Table \ref{tab:tab2}  summarizes the results for all the attack methods. In terms of attack outcome, the proposed TDP-CP achieves the best performance and the improvement is significant. For example, compared with the best baseline, the proposed ’s MRR value increases by over 20\% on average. Among the baseline methods, \textit{Random} simply makes the words of the target news replace the words of the history clicked news without actually creating any new pattern that favors the news recommendation of the target news. Therefore, \textit{Random} has the worst performance and even makes the MRR value worse. \textit{Effective} ensures that the news we modify each time is the most effective one. However, it might involve two perturbations with duplicate promoting effects on the target news, that is to say, using one of them can achieve the same promotion as using both.  Thus, it leads to a waste of budget, but cannot get a satisfactory performance. Compared with these baselines, the proposed TDP-CP takes advantages of the MDP manner with feedback from the influence estimator to train an attack agent. The learned agent is capable of creating more complex patterns to achieve the data poisoning goal.
\begin{table}[h]
\centering
\caption{Overall performance}\label{tab:tab2}
\begin{tabular}{|l|l|l|l|}
\hline
Method & DeepFM & NAML & NRMS  \\\hline \hline
None & 0.187$\pm$0.026 & 0.152$\pm$0.028 & 0.163$\pm$0.029 \\
Random & 0.193$\pm$0.031 & 0.138$\pm$0.024 & 0.176$\pm$0.027\\
Effective & 0.253$\pm$0.023 & 0.240$\pm$0.021 & 0.252$\pm$0.024 \\
TDP-CP & \textbf{0.291$\pm$0.025} & \textbf{0.290$\pm$0.023} & \textbf{0.328$\pm$0.020}\\
\hline
\end{tabular}
\end{table}

\subsection{Ablation Study}
\subsubsection{Two-Stage Hierarchical Structure}
We verify the advantage of our policy network with a two-stage hierarchical structure (Sec.~\ref{subsec:opt}) with two variants. One is without hierarchical structure (\textit{non-HS}), and the other one is not to align nodes based on effectiveness but construct the tree with the random ordering of the leaf nodes (\textit{Random-HS}).  In order to show the performance, we use the MRR without any attack as a base, then calculate the MRR changing along with the training process.

We illustrate the results evaluated on NRMS as the online recommendation system in Fig.~\ref{fig:hs}, which shows that our proposed structure is far better than the rest.    For the policy network without hierarchical structure whose search space is $\mathrm{O}(|\mathcal{A}^r|\times L^2)$ where $L$ is the average length of the title, two-stage hierarchical structure decreases the search space to $\mathrm{O}(\log|\mathcal{A}^r| + 2\log L)$, as a result, two-stage hierarchical structure achieves a better performance with the same training steps.  For the hierarchical structure with random node arrangement, as the local structure cannot help to shrink the space because of effectiveness gaps, although the search space is reduced, the selected policy is not optimal.

\begin{figure}[h]
    \centering
    \includegraphics[width=8cm]{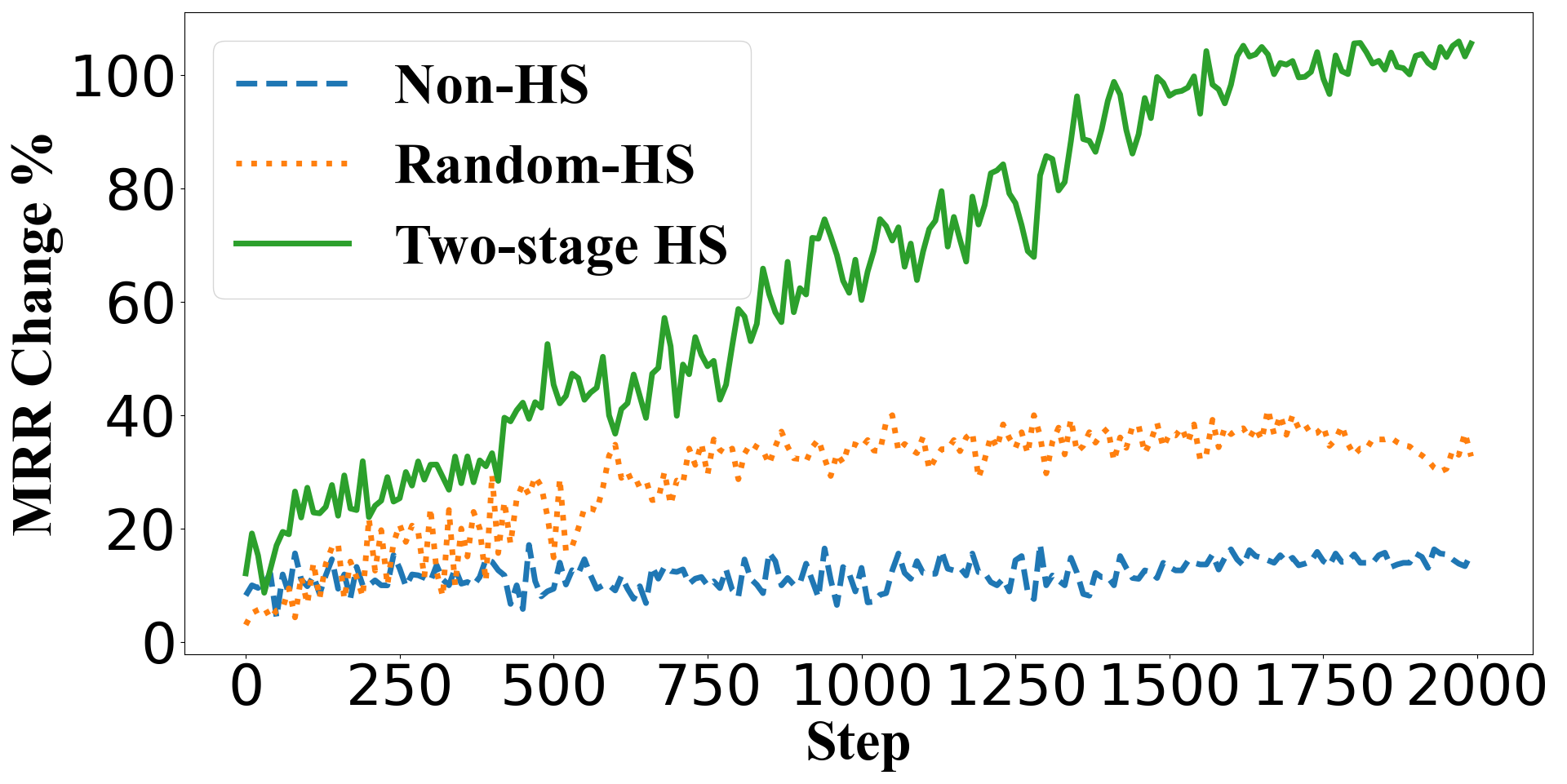}
    \caption{Advantages of Two-Stage Hierarchical Structure.}
    \label{fig:hs}
\end{figure}

\subsubsection{Influence Estimation} 
We verify the effect of the influence estimator (Sec.~\ref{subsec:infest}) via estimation accuracy and executive time to calculate the reward. As shown in the Tab. \ref{tab:Influ}, the estimated reward achieves an accuracy of about 98.7\% relative to the real training on average of three recommendation systems. The comparison of the average duration of a feedback given by the environment under the algorithm is also shown in Tab. \ref{tab:Influ}. We can clearly observe a nearly 50 times improvement in the efficiency of the introduction of the influence estimation function.

\begin{table}[h]
\centering
\caption{Influence estimation}\label{tab:Influ}
\begin{tabular}{|c||l||l|l|}
\hline
News Rec & Estimation & \multicolumn{2}{c}{Executive Time (s) } \vline \\ \cline{3-4}
 Sys&  Accuracy   & w/o Inf Est & with Inf Est \\\hline \hline
DeepFM & 99.4\% & 53.49 & 1.37\\
NAML & 98.4\% & 60.28 & 1.43\\
NRMS & 98.3\% & 58.33 & 1.42\\
\hline
\end{tabular}
\end{table}

\section{Conclusion and future work}
In this study, we propose a novel targeted data poisoning approach for a news recommendation system of perturbing the news content.  We exploit the power of deep reinforcement learning to solve it effectively with a two-stage hierarchical structure to reduce the search space and an influence estimator to speed up reward calculation.  We conduct extensive experiments on three representative recommendation algorithms and the news recommendation benchmark dataset. The results show that our TDP-CP can increase the rank of target news successfully with a limited exposure budget.  

The attack considered in the paper is a targeted, risk-constrained attack. A natural future direction is to study whether it is possible to learn a general attack model which is not specific to a target. Secondly, the perturb operator we used is title word replacement, but to generate the perturb content is a promising direction to reduce the exposure. Finally, it is important to develop defense mechanisms against such attacks with high efficiency and low overhead for practice systems.

\newpage
\bibliography{main}

\newpage
\section{Appendix}
\subsection{More Experiment Results}
The experiments are implemented by Python 3.6 and conducted on a server with four core Nvidia GeForce RTX 2080 Ti(each run requires two cores) and 256GB RAM of memory.

\subsubsection{The Effect of the Cost Function.} 

As shown in Eq.~\ref{eqn:grisk}, we designed the risk function to measure the risk of our perturbation. The risk function imposes two constraints on perturbation, the frequency of the news item we are perturbing and the content similarity before and after perturbed. Intuitively, in order to make our perturbation less likely to be discovered, we want the exposure of the modified news to be relatively low and the content we change to be as similar as possible.

\begin{figure}[h]
    \centering
    \includegraphics[width=8cm]{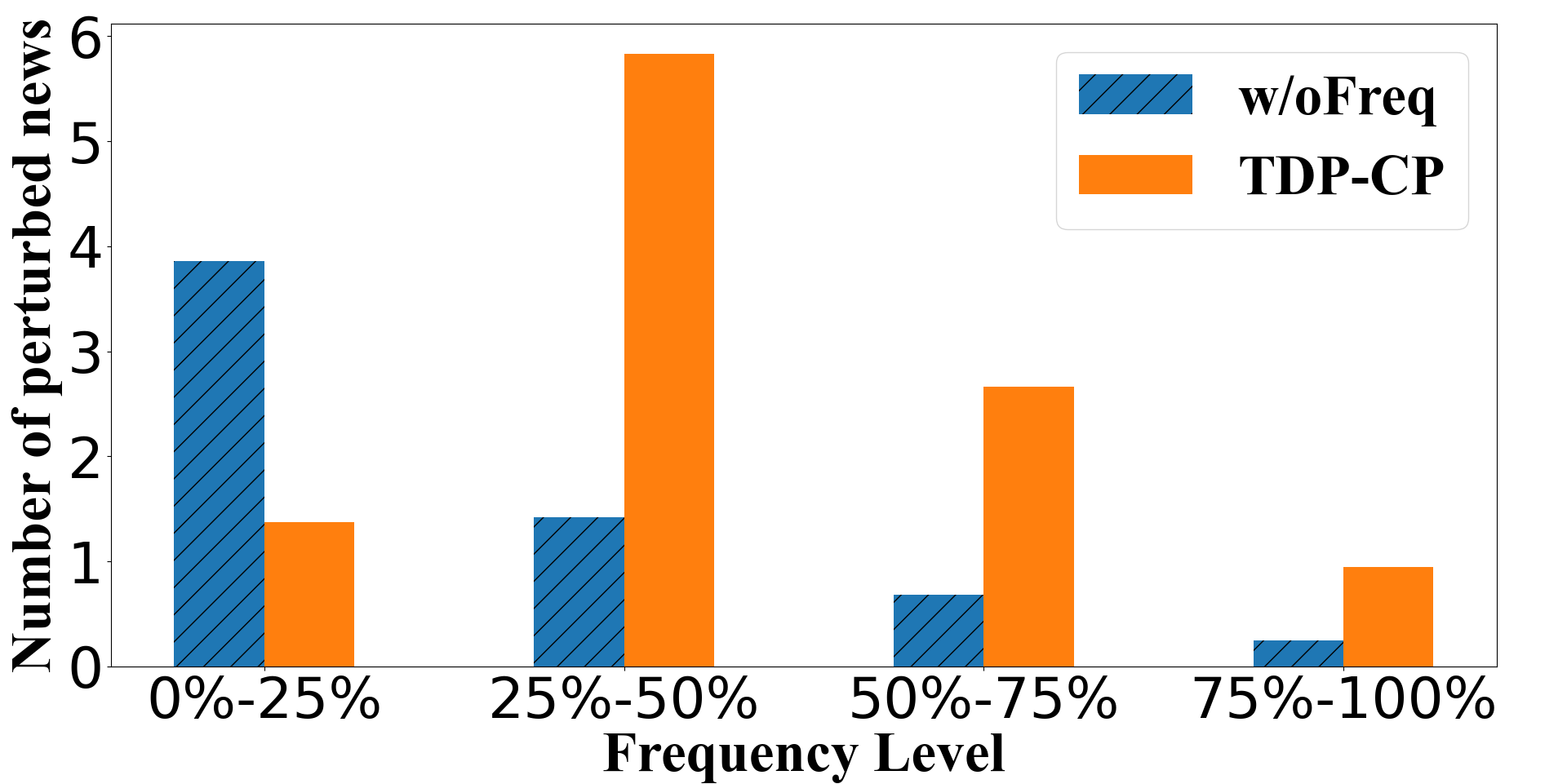}
    \caption{Advantages of the Frequency Cost Function}
    \label{Pcost}
\end{figure}

In our experiments, we verified the practical effects of the two parts of the risk function. First, as shown in the Fig. \ref{Pcost}, this left part is the validation of the frequency risk function. We designed an experiment with the frequency constraint removed from the risk function. With the same performance guaranteed(80\% improvement over None), we conduct experiments. Specifically, we used the number of clicks on the news as a frequency metric and divided them into four classes (0-25\%, 25\%-50\%, 50\%-75\%, 75\%-100\%) according to their ranking, where 0-25\% is the top 25\% most popular candidates. In the Fig. \ref{Pcost}, we can clearly see that with the frequency constraint (blue) compared to without it (orange), the distribution of modified news is significantly shifted back, tending to change news that is relatively less popular. In addition to this, we also find that the total number of perturbations to the news increases for agents with frequency constraints in order to ensure that similar results are achieved.

\begin{figure}[h]
    \centering
    \includegraphics[width=8cm]{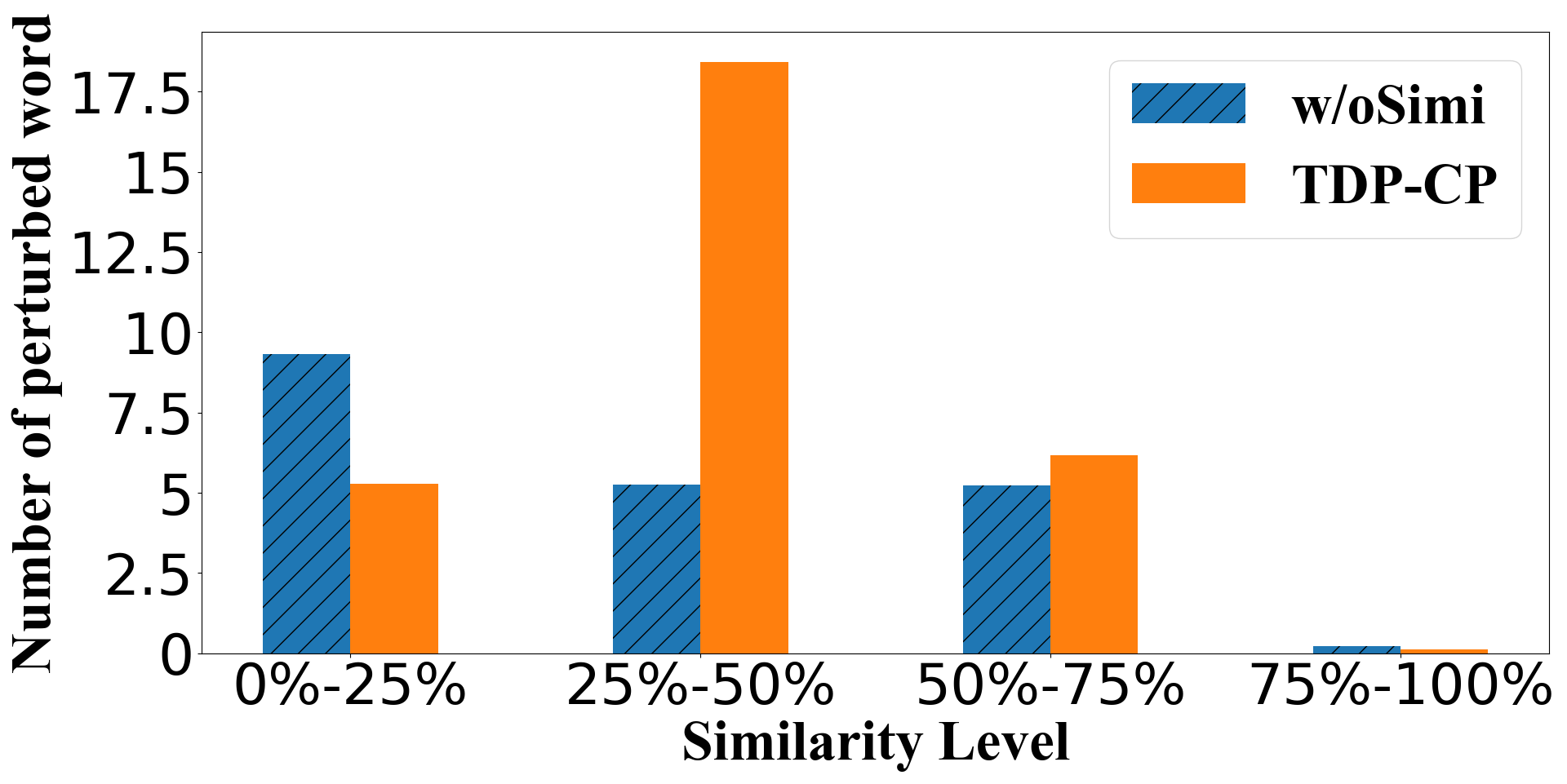}
    \caption{Advantages of the Similarity Cost Function}
    \label{fig:Scost}
\end{figure}

Similarly, we validated the validity of another similarity constraint in the risk function. As shown in the Fig. \ref{fig:Scost}, we classify the similarity of the replaced words into four classes, where 0-25\% indicates the least similar class. The percentage of each part in the total number of modified words is investigated. The effect is evident that by adding the similarity constraint, the agent shifts the modification scheme and is more inclined to make before and after perturbation relatively similar.

In addition, we find very similar modifications, corresponding to the 75\%-100\% part of the graph, always out of a minimum state. The most plausible explanation for this phenomenon would be that our modifications, to achieve their purpose, must choose some relatively less identical words to replace in order to have a different impact. The extreme case of an exact agreement would be equivalent to not making any modification, so the similarity constraint makes a trade-off between risk and effect.

\subsubsection{Deviation of the recommended simulation.} 

\begin{table}[h]
\centering
\caption{Deviation of the recommended simulation}\label{tab:recoSimu}
\begin{tabular}{|l|l|l|l|}
\hline
Method & DeepFM & NAML & NRMS  \\\hline \hline
None & -0.002 & +0.004 & \textbf{+0.005}\\
Random & +0.003 & -0.001 & +0.002\\
Cost-effective & -0.003 & +0.002 & +0.002\\
TDP-FN & -0.002 & +0.001 & +0.003\\
\hline
\end{tabular}
\end{table}

In our experiments, we use an offline model to simulate an online news recommendation system. To verify the effect of our simulator on the final attack effect. We replace the simulator in our experiments with the real recommendation system to be attacked. We show the changes in the results after adding the simulator in the Table \ref{tab:recoSimu}. We can see that the effect of the simulator on the effectiveness of the attack is small, with the largest change being 0.05. In addition, we also find that our results are slightly worse when we attack a generic online recommendation model (DeepFM) using the model learned in the current simulation environment. The reason for this is that DeepFM is worse at extracting information at the word level so that some perturbations do not achieve the desired effect.

\end{document}